\documentclass[conference]{IEEEtran}
\IEEEoverridecommandlockouts
\usepackage{cite}
\usepackage{amsmath,amssymb,amsfonts}
\usepackage{algorithmic}
\usepackage{graphicx}
\usepackage{textcomp}
\usepackage{booktabs}
\usepackage{graphicx}
\usepackage{multirow}
\usepackage{hyperref}
\usepackage{url}
\usepackage[table,xcdraw]{xcolor}

\usepackage{tikz, url, hyperref, eso-pic}
\hypersetup{
    colorlinks=true,
    linkcolor=blue,      
    citecolor=red,       
    filecolor=magenta,   
    urlcolor=blue        
}

\newcommand{\addpublisheddetails}[3]{%
    \AddToShipoutPictureBG*{%
        \AtPageUpperLeft{%
            \begin{tikzpicture}[remember picture,overlay]
                \node[anchor=north, fill=yellow!30, text width=\dimexpr\paperwidth-2cm, align=left, font=\sffamily\footnotesize, inner sep=8pt] 
                at ([xshift=0cm, yshift=0cm]current page.north) {%
                    \baselineskip=0.95\baselineskip
                    \textbf{Citation:} #2 \\[0.4em]
                    \textbf{Access:} \url{#3} - \textbf{Venue:} ICASSP 2025
                };
            \end{tikzpicture}
        }
    }
}
\addpublisheddetails{ICASSP 2025}
{La Quatra, M., Orozco-Arroyave, J. R., Siniscalchi, M. S., ``Bilingual Dual-Head Deep Model for Parkinson’s Disease Detection from Speech,'' ICASSP, 2025, pp. 1-5, doi:10.1109/ICASSP49660.2025.10889445}
{https://doi.org/10.1109/ICASSP49660.2025.10889445}

\definecolor{ewa_blue}{HTML}{DAE8FC}
\definecolor{pcgita_yellow}{HTML}{FFFFC7}

\def\BibTeX{{\rm B\kern-.05em{\sc i\kern-.025em b}\kern-.08em
    T\kern-.1667em\lower.7ex\hbox{E}\kern-.125emX}}

\begin{document}

\title{Bilingual Dual-Head Deep Model for Parkinson's Disease Detection from Speech %
\thanks{
This work was partially supported by the "D.A.R.E. - Digital Lifelong Prevention" project (PNC0000002, CUP: B53C22006450001) and 
Miur Prin 2022 - SHAPE-AD, CUP: {J53D23007240008}.
}
}

\author{\IEEEauthorblockN{Moreno La Quatra}
\IEEEauthorblockA{\textit{Kore University of Enna} \\
Enna, Italy \\
moreno.laquatra@unikore.it}
\and
\IEEEauthorblockN{{Juan Rafael }{Orozco-Arroyave}}
\IEEEauthorblockA{\textit{University of Antioquia} \\
Medellin, Colombia \\
rafael.orozco@udea.edu.co}
\and
\IEEEauthorblockN{Marco Sabato Siniscalchi}
\IEEEauthorblockA{\textit{UniPA / NTNU} \\
Palermo, Italy / Trondheim, Norway \\
sabatomarco.siniscalchi@unipa.it}
}

\maketitle

\begin{abstract}

This work aims to tackle the Parkinson’s disease (PD) detection problem from the speech signal in a bilingual setting  by proposing an ad-hoc dual-head deep neural architecture for type-based binary classification. One head is specialized for  diadochokinetic patterns. %
The other head looks for natural speech patterns present in continuous spoken utterances. Only one of the two heads is operative accordingly to the nature of the input. 
Speech representations are extracted from self-supervised learning (SSL) models and wavelet transforms. 
Adaptive layers, convolutional bottlenecks, and contrastive learning are exploited to reduce variations across languages. 
Our solution is assessed against two distinct datasets, EWA-DB, and PC-GITA, which cover  Slovak and Spanish languages, respectively.
Results indicate that conventional models trained on a single language dataset struggle with cross-linguistic generalization, and naive combinations of datasets are suboptimal. In contrast, our model improves generalization on both languages, simultaneously.

\end{abstract}

\begin{IEEEkeywords}
    Parkinson's Disease Detection, Speech Processing, Self-Supervised Learning
\end{IEEEkeywords}

\section{Introduction}
\label{sec:introduction}

The detection of Parkinson's disease (PD) from the speech signal has gained significant attention in recent years due to the non-invasive nature of speech recordings, and the potential for early diagnosis, e.g., \cite{tsanas2012novel, orozco2016automatic, Vasquez2018, perez2019natural, vasquez2021transfer, sonawane2021speech, Narendra2021, garcia2022detecting, quan2022end, favaro2023multilingual,Reddy2023}. PD detection from speech often relies on analyzing various types of speech tasks, including spoken or read text exercises, diadochokinetic (DDK) exercises (e.g., the rapid repetition of syllables like "pa-ta-ka"), isolated words, and sustained vowels.  Spoken or read text exercises provide insights into natural speech patterns, while DDK exercises are valuable for assessing motor control and articulatory precision, which are frequently impaired in PD \cite{bocklet2013automatic}.  Isolated words and sustained vowels offer additional information on articulation and phonation, respectively.  We here focus on the most widely available tasks across different datasets, namely DDK, and read/spontaneous speech. 
These tasks highlight complementary speech features crucial for detecting PD and providing a comprehensive assessment of speech impairments related to the disease.

Recent studies have shown that deep learning models can achieve high accuracy in PD detection from speech data \cite{la2024exploiting}, with the potential to outperform traditional pipeline methods based on handcrafted features \cite{escobargrisales23_interspeech}.
However, the generalization of these models across different languages remains a significant challenge due to variations in linguistic and recording conditions \cite{kovac2021multilingual}.
This issue is particularly relevant in the context of PD detection, where the linguistic background of the speakers can significantly impact the performance of the models. 
The primary challenge in cross-linguistic generalization arises from the inherent variability in speech characteristics across different languages and speakers. 
Models trained on data from a single language often struggle to maintain high performance when applied to data from another language, leading to a significant drop in accuracy and robustness \cite{kovac2021multilingual}. 
Moreover, straightforward approaches, such as combining multiple datasets into a single training set, have shown limited success due to the complex interactions between linguistic features and PD-specific speech characteristics.

To address these challenges, we propose a novel dual-head deep  architecture that incorporates task-specific branches to enhance the detection of Parkinson's disease from speech across different languages. 
Unlike existing models, which often use a single path to process all types of speech data, our model employs a shared backbone for feature extraction, followed by two separate branches for task-specific PD detection. Specifically, one head is specialized on the DDK task; whereas, the other head focuses on continuous speech. This dual-head architecture allows the model to specialize in different speech tasks, effectively capturing unique patterns and nuances specific to each type of speech data.
The backbone extracts speech representation leveraging an SSL model \cite{baevski2020wav2vec,chen2022wavlm,hsu2021hubert}, and the wavelet transform \cite{Mallat2008}.
To improve cross-linguistic generalization, an adaptive normalization technique that adjusts feature statistics dynamically across different datasets is employed, thereby reducing domain shifts \cite{huang2017adain}. Moreover, convolutional bottleneck layers are used to compress and expand the feature space, enhancing the model's ability to focus on relevant speech patterns while filtering out irrelevant information. Finally, contrastive learning is employed to boost discriminative capabilities across languages.
Experimental evidence  on EWA-DB (Slovak) \cite{rusko2023ewa} and PC-GITA (Spanish) \cite{orozco2014new, karan2021non} demonstrate 
that our dual-head architecture combining both SSL- and Wavelet-based speech representations can handle both languages, outperforming baseline models that fail to generalize. 
In contrast, our method effectively improves generalization across languages. 

\section{Related Work}
\label{sec:related_work}

Despite end-to-end deep models have proven effective for automatic PD screening from speech \cite{la2024exploiting}, these models face challenges in generalization across different languages \cite{ibarra2023towards}. 
Domain adaptation techniques mitigate the impact of domain shifts caused by linguistic diversity and varying recording conditions, showing promise in maintaining performance despite the variability inherent in impaired speech \cite{woszczyk20_interspeech, wang21u_interspeech}. In a cross-language setting, \cite{rios2024automatic} proposed a domain adaptation technique using Gaussian mixture models  to discriminate PD from essential tremor (ET) across different languages, achieving notable results in both bi-class and tri-class classification tasks. In \cite{Orozco2016},  the energy content of the unvoiced sounds on different speech recordings for Spanish, German and Czech were computed and both in-language and cross-language results were reported. Nonetheless, multi-lingual experiments were not reported and different models were built for each language.
Therefore, multilingual generalization remains a critical challenge in PD detection research \cite{ibarra2023towards}.

In recent research, self-supervised learning (SSL) \cite{hsu2021hubert, baevski2020wav2vec, chen2022wavlm} models have reported state-of-the-art  PD detection results on both standard and extended Spanish PC-GITA datasets \cite{la2024exploiting}, due to their ability to learn robust feature representations from large unlabeled datasets. Nonetheless, results on cross-lingual and multilingual setups are unavailable to the best of the authors' knowledge.
In \cite{torghabeh2023enhancing}, wavelet transforms have been used to extract multi-scale features from speech, which were shown to capture fine-grained temporal patterns indicative of Parkinson's disease. Yet, the experiments were carried out considering only an in-language experimental setup.

\section{Proposed Method}
\label{sec:proposed_method}

\begin{figure}[htbp]
    \centering
    \includegraphics[width=\columnwidth]{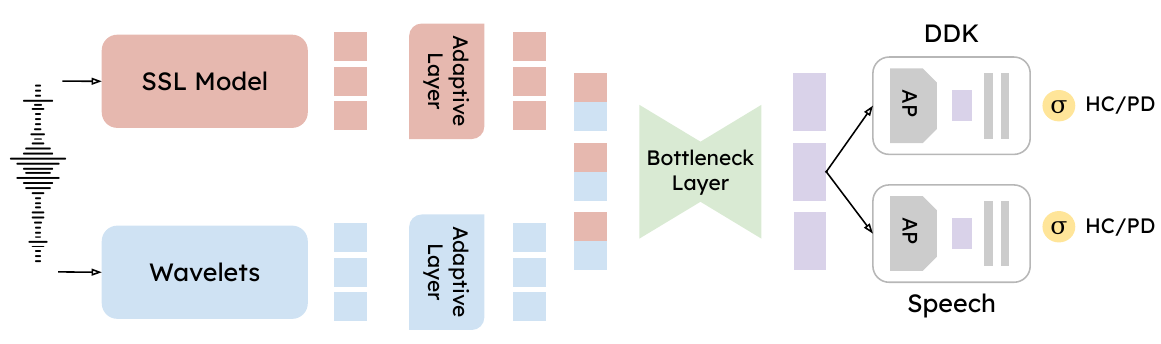}
    \caption{Proposed model architecture. Each branch includes an attention pooling layer (AP) and two linear classification layers with ReLU. HC refers to Healthy Control, PD to Parkinson's Disease, and \(\sigma\) is the sigmoid activation.}
    \label{fig:model_architecture}
\end{figure}

The proposed model architecture,  illustrated in Figure \ref{fig:model_architecture}, aims to improve cross-language generalization for PD detection by leveraging a combination of SSL features, wavelet transforms, and specific architectural techniques, namely adaptive layer, convolutional bottleneck, and contrastive learning. 
The model consists of several key components designed to address the challenges of cross-language generalization and robust feature extraction.

\subsection{Type-Based Classifiers}

Our model employs a dual-head architecture to handle the distinct characteristics of different speech tasks, specifically Diadochokinetic (DDK) exercises and read/spontaneous continuous speech. 
This setup includes separate branches for each task, enabling specialized processing. 
The shared backbone extracts features $\mathbf{z}_{DDK}$ and $\mathbf{z}_{speech}$, which are then fed into task-specific classifiers:
\begin{equation}
    \mathbf{y}_{DDK} = f_{DDK}(\mathbf{z}_{DDK}), \quad \mathbf{y}_{speech} = f_{speech}(\mathbf{z}_{speech}),
\end{equation}
where $f_{DDK}$ and $f_{speech}$ are classifiers for the DDK and speech tasks, respectively. 
Each classifier includes an attention pooling (AP) layer, which summarizes the input sequence into a single vector by weighting frames based on their importance with learnable parameters. 
This vector then passes through two fully connected layers with ReLU activation, enabling the model to capture distinct patterns for each type of speech while utilizing common features across tasks.

\subsection{Wavelet Integration}

To enhance feature representation, we integrate wavelet-based features with SSL features at a frame level. 
The input speech signal, $\mathbf{x}$, is divided into overlapping frames to capture localized temporal dynamics. 
For each frame, SSL features $\mathbf{z}_{SSL}$ are extracted using a pre-trained SSL model. 
Simultaneously, wavelet decomposition is applied to each frame to extract detailed coefficients, resulting in frame-level wavelet features $\mathbf{z}_{wavelet}$.
To ensure consistency in feature scales, both SSL and wavelet representation are normalized using Layer Normalization (LN):
\begin{equation}
    \mathbf{z}_{SSL}^{\text{norm}} = \text{LN}(\mathbf{z}_{SSL}), \quad \mathbf{z}_{wavelet}^{\text{norm}} = \text{LN}(\mathbf{z}_{wavelet}).
\end{equation}
The normalized features for each frame are concatenated along the feature dimension to form a sequence of vectors:
\begin{equation}
    \mathbf{z}_{concat} = \text{Concat}(\mathbf{z}_{SSL}^{\text{norm}}, \mathbf{z}_{wavelet}^{\text{norm}}).
\end{equation}
This sequence, $\mathbf{z}_{concat}$, combines high-level semantic features from SSL with detailed temporal features from wavelet analysis, improving the model's capacity to detect PD-related patterns in speech, as shown later in our experiments.

\subsection{Adaptive Layers}

To handle variations in speech data across languages, we utilize adaptive layers inspired by AdaIN \cite{huang2017adain}. 
These layers adjust feature statistics to minimize domain shifts caused by linguistic differences.
The adaptive layer first normalizes the input features $\mathbf{x}$ using their mean and standard deviation to obtains $\mathbf{z}_{\text{norm}}$.

The normalized features are then modulated using language-conditioned scaling and shifting parameters, $\gamma$ and $\beta$, derived from learnable, domain-specific embeddings:
\begin{equation}
    \gamma = g(\mathbf{e}_{\text{lang}}), \quad \beta = h(\mathbf{e}_{\text{lang}}),
\end{equation}
with $\mathbf{e}_{\text{lang}}$ representing the language-specific embedding vector, and $g(\cdot)$ and $h(\cdot)$ being learnable functions that map these embeddings to the modulation parameters. 
The final output of the adaptive layer (AL) is computed as:
\begin{equation}
    \mathbf{z}_{\text{AL}} = \gamma \cdot \mathbf{z}_{\text{norm}} + \beta.
\end{equation}
The feature distributions are dynamically adjusted according to the linguistic context via the adaptive mechanism, reducing domain shifts and enhancing cross-language generalization.

\vspace{-1mm}
\subsection{Bottleneck Layers}

To refine feature representations and emphasize relevant speech characteristics, we use CNN-based bottleneck layers. 
These layers perform compression and expansion in the feature space, allowing the model to focus on key speech features while reducing irrelevant information.
First, the input features $\mathbf{z}$ are compressed using a convolutional layer with a ReLU activation, reducing the dimensionality of the feature space:
\begin{equation}
    \mathbf{z}_{\text{compressed}} = \text{ReLU}(W_1 \ast \mathbf{z}),
\end{equation}
where $W_1$ is a convolutional filter, and $\ast$ denotes the convolution operation. 
The compressed features are then expanded back to their original dimensionality using a second convolutional layer:
\begin{equation}
    \mathbf{z}_{\text{expanded}} = W_2 \ast \mathbf{z}_{\text{compressed}},
\end{equation}
with $W_2$ serving as the filter for feature expansion. 
A sigmoid activation is then applied to selectively retain important features:
\begin{equation}
    \mathbf{z}_{\text{bottleneck}} = \sigma(\mathbf{z}_{\text{expanded}}) \cdot \mathbf{z} %
\end{equation}
This operation should allow the model to enhance key speech features while filtering out non-relevant temporal variations.
Finally,  a residual connection is used, and  bottleneck features are summed with the original $\mathbf{z}$.

\vspace{-1mm}
\subsection{Contrastive Learning}

To enhance the model's discriminative capability, we employ a contrastive learning framework that encourages the separation of different classes in the feature space. 
The model is trained to increase the similarity of positive pairs (same class) while decreasing the similarity of negative pairs (different classes). 
The contrastive loss function is formulated as:
\begin{equation}
    \mathcal{L}_{\text{contrastive}} = (d_{\text{pos}} - m_{\text{pos}}) + (m_{\text{neg}} - d_{\text{neg}}),
\end{equation}
where $d_{\text{pos}}$ and $d_{\text{neg}}$ are the distances between positive and negative pairs, respectively, and $m_{\text{pos}}$ and $m_{\text{neg}}$ are margin thresholds that define the min and max allowable distances for these pairs. 
To target the most informative samples, we implement a custom miner that identifies the hardest positive and negative pairs. 
It selects positive pairs with the greatest distance $d_{\text{pos}}$ and negative pairs with the smallest distance $d_{\text{neg}}$:
\begin{equation}
    \{i^+, j^+\} = \arg \max_{i, j} \, d_{\text{pos}}, \quad \{i^-, j^-\} = \arg \min_{i, j} \, d_{\text{neg}},
\end{equation}
where $\{i^+, j^+\}$ and $\{i^-, j^-\}$ represent the indices of the hardest positive and negative pairs, respectively. 
By focusing on these challenging examples, the model refines its feature representation, enhancing its ability to distinguish subtle speech patterns that are indicative of PD.

\section{Experiments}
\label{sec:experiments}

To evaluate the effectiveness of our proposed model for PD detection from speech, we conducted experiments using two distinct datasets: EWA-DB \cite{rusko2023ewa} and PC-GITA \cite{orozco2014new}. 

\vspace{-1mm}
\subsection{Datasets}

\noindent
\textit{EWA-DB:} 
This dataset includes recordings from Slovak speakers, comprising 863 healthy controls (HC) and 95 PD patients, with no overlap between speakers in the training, validation, and test sets (70\%, 10\%, 20\% split). 
Each speaker provided five spontaneous speech recordings and performed a Diadochokinetic (DDK) task involving "pataka" repetitions. 
Due to the dataset's imbalance, weighted sampling was employed to ensure balanced training.

\vspace{1mm}
\noindent
\textit{PC-GITA:} 
The PC-GITA dataset consists of Spanish speech recordings, split into standard (s-PC-GITA) and extended (e-PC-GITA) versions \cite{karan2021non}. 
s-PC-GITA contains clean recordings used for training and validation (20\% validation split), while e-PC-GITA, containing real-world recording conditions, served as test set. 
The complete dataset includes 140 speakers (70 PD patients, 70 HC). We cover a range of tasks including read text, monologue, sentence pronunciation, and several DDK tasks (e.g., including different syllable repetitions).

\noindent
To ensure consistency, following previous work \cite{la2024exploiting}, both the e-PC-GITA test set and all EWA-DB splits were preprocessed using voice activity detection (VAD) \cite{tan2020rvad}, speech dereverberation \cite{welker2022speech}, and speech denoising \cite{lu23e_interspeech}.
These steps ensured uniform data preparation, enabling fair performance comparisons across different linguistic and recording settings.

\vspace{-1mm}
\subsection{Experimental Setup and Metrics}

Three different setups were used to assess our solution:
(1) \textit{In-Language}, training and testing on the same dataset (EWA-DB or PC-GITA); 
(2) \textit{Cross-Language}, training on one dataset and testing on the other to assess generalization; 
(3) \textit{BiLingual}, training on both datasets to explore the impact of combining diverse collections.

\vspace{1mm}
\noindent
\textit{Evaluation Metrics:}
Model performance was assessed using four key metrics: 
\textit{Accuracy} (the proportion of correctly classified samples), 
\textit{F1 Score} (the harmonic mean of precision and recall), 
\textit{Sensitivity} (the true positive rate), and 
\textit{Specificity} (the true negative rate). 

\vspace{1mm}
\noindent
\textit{Training Procedure:}
All models were trained on one NVIDIA A100 GPU using the AdamW optimizer \cite{loshchilov2018decoupled} with a batch size of 64.
The maximum learning rate was set to $10^{-4}$, with a warm-up ratio of 0.1 and linear decay.
Training was conducted for 20 epochs with early stopping based on validation F1 score.
Based on previous findings \cite{la2024exploiting}, WavLM base \cite{chen2022wavlm} was selected as the SSL model for feature extraction\footnote{Code available at: \href{https://github.com/MorenoLaQuatra/BDHPD}{\texttt{https://github.com/MorenoLaQuatra/BDHPD}}}.

\begin{table}[]
\centering
\caption{Model performance on \colorbox{ewa_blue}{EWA-DB} and \colorbox{pcgita_yellow}{PC-GITA} datasets for different training configurations. 
}
\vspace{-2mm}

\label{tab:performance_comparison}
\resizebox{\columnwidth}{!}{%
\begin{tabular}{@{}cccccc@{}}
\toprule
\begin{tabular}[c]{@{}c@{}}Training \\ Dataset\end{tabular} &
  Test Dataset &
  Accuracy &
  F1 Score &
  Sensitivity &
  Specificity \\ \midrule
 &
  \cellcolor[HTML]{DAE8FC}EWA-DB &
  \cellcolor[HTML]{DAE8FC}83.59 &
  \cellcolor[HTML]{DAE8FC}\textbf{70.28} &
  \cellcolor[HTML]{DAE8FC}69.57 &
  \cellcolor[HTML]{DAE8FC}85.5 \\
\multirow{-2}{*}{\begin{tabular}[c]{@{}c@{}}In-\\ Language\end{tabular}} &
  \cellcolor[HTML]{FFFFC7}PC-GITA &
  \cellcolor[HTML]{FFFFC7}90.00 &
  \cellcolor[HTML]{FFFFC7}90.00 &
  \cellcolor[HTML]{FFFFC7}91.67 &
  \cellcolor[HTML]{FFFFC7}\textbf{88.33} \\ \midrule
 &
  \cellcolor[HTML]{DAE8FC}EWA-DB &
  \cellcolor[HTML]{DAE8FC}23.96 &
  \cellcolor[HTML]{DAE8FC}23.91 &
  \cellcolor[HTML]{DAE8FC}\textbf{89.13} &
  \cellcolor[HTML]{DAE8FC}15.09 \\
\multirow{-2}{*}{\begin{tabular}[c]{@{}c@{}}Cross-\\ Language\end{tabular}} &
  \cellcolor[HTML]{FFFFC7}PC-GITA &
  \cellcolor[HTML]{FFFFC7}65 &
  \cellcolor[HTML]{FFFFC7}63.54 &
  \cellcolor[HTML]{FFFFC7}45 &
  \cellcolor[HTML]{FFFFC7}85 \\ \midrule
 &
  \cellcolor[HTML]{DAE8FC}EWA-DB &
  \cellcolor[HTML]{DAE8FC}75.78 &
  \cellcolor[HTML]{DAE8FC}60.74 &
  \cellcolor[HTML]{DAE8FC}57.97 &
  \cellcolor[HTML]{DAE8FC}78.21 \\
\multirow{-2}{*}{\begin{tabular}[c]{@{}c@{}}Bilingual\\ (Baseline)\end{tabular}} &
  \cellcolor[HTML]{FFFFC7}PC-GITA &
  \cellcolor[HTML]{FFFFC7}78.33 &
  \cellcolor[HTML]{FFFFC7}78.33 &
  \cellcolor[HTML]{FFFFC7}80.0 &
  \cellcolor[HTML]{FFFFC7}76.67 \\ \midrule
 &
  \cellcolor[HTML]{DAE8FC}EWA-DB &
  \cellcolor[HTML]{DAE8FC}\textbf{84.72} &
  \cellcolor[HTML]{DAE8FC}69.03 &
  \cellcolor[HTML]{DAE8FC}56.52 &
  \cellcolor[HTML]{DAE8FC}\textbf{88.56} \\
\multirow{-2}{*}{\begin{tabular}[c]{@{}c@{}}Bilingual\\ (Proposed)\end{tabular}} &
  \cellcolor[HTML]{FFFFC7}PC-GITA &
  \cellcolor[HTML]{FFFFC7}\textbf{90.83} &
  \cellcolor[HTML]{FFFFC7}\textbf{90.83} &
  \cellcolor[HTML]{FFFFC7}\textbf{93.33} &
  \cellcolor[HTML]{FFFFC7}\textbf{88.33} \\ \bottomrule
\end{tabular}%
}
\end{table}

\begin{figure}[htbp]
\centering
\includegraphics[width=\columnwidth]{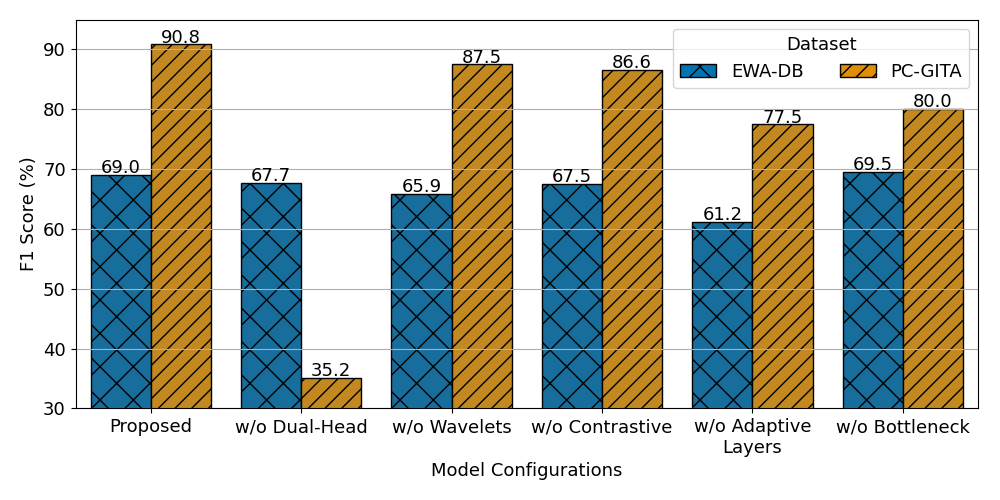}
\vspace{-5mm}
\caption{Ablation study results showing the impact of removing individual components on the F1 score for EWA-DB and PC-GITA datasets.}
\vspace{-5mm}

\label{fig:ablation_study}
\end{figure}

\begin{figure}[htbp]
    \centering
    \begin{minipage}[b]{0.49\columnwidth}
        \centering
        \includegraphics[width=\columnwidth]{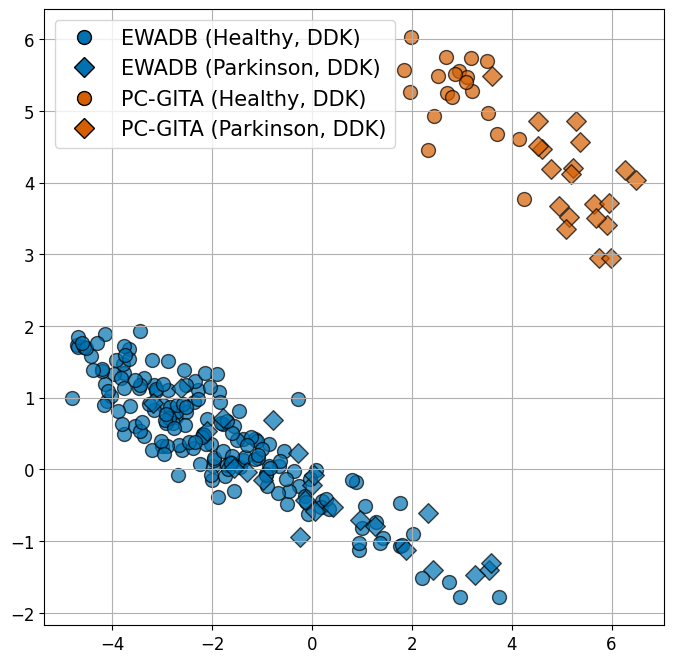}
        \small\textit{(a) Without Dual-Head, language-discriminative features.}%
        \label{fig:no_tbc}
    \end{minipage}
    \hfill
    \begin{minipage}[b]{0.49\columnwidth}
        \centering
        \includegraphics[width=\columnwidth]{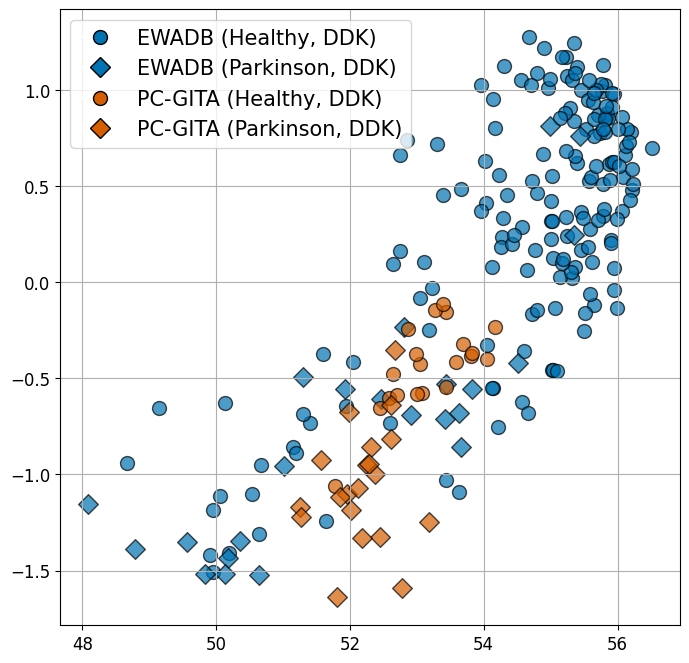}
        \small\textit{(b) With Dual-Head, language-shared features.}%
        \label{fig:tbc}
    \end{minipage}
    \vspace{-5mm}
    \caption{t-SNE visualization of DDK task embeddings.}
    \label{fig:tsne_comparison}
\end{figure}

\vspace{-1mm}
\subsection{Results and Analysis}
\label{subsec:results_analysis}

Table \ref{tab:performance_comparison} summarizes the performance of the proposed model across different training configurations.

\noindent
\textit{In-Language Experiments:}
The model performed well when trained and tested on the same dataset, achieving 83.59\% accuracy and 70.28 F1 score on EWA-DB, and 90\% accuracy and 90.00 F1 score on PC-GITA. 
This demonstrates the model's capability to capture PD-specific speech patterns within a single linguistic context.

\vspace{1mm}
\noindent
\textit{Cross-Language Experiments:}
A significant performance drop occurred when the model was tested on a dataset different from the training one (e.g., 23.96\% accuracy, 23.91 F1 score for PC-GITA $\rightarrow$ EWA-DB), highlighting the challenge of cross-language generalization due to variability in speech characteristics.
This drop highlights the challenges associated with cross-language generalization.%

\vspace{1mm}
\noindent
\textit{Bilingual Experiments:}
Training on both datasets improved generalization, but the baseline model, 
which does not employs a dual-head architecture, 
Adaptive Layer, Convolutional Bottleneck, and Contrastive Learning attains suboptimal results (75.78\% accuracy, 60.74 F1 score on EWA-DB; 78.33\% accuracy, 78.33 F1 score on PC-GITA). 
In contrast, the proposed model, Bilingual (Proposed), improved performance, achieving 84.72\% accuracy and 69.03 F1 score on EWA-DB, and 90.83\% accuracy and 90.83 F1 score on PC-GITA, demonstrating the effectiveness of our approach in handling two languages with a single model.
The low sensitivity on EWA-DB likely results from the smaller PD sample size, causing overfitting in specificity due to the over-representation of HC and under-representation of PD acoustic patterns.

\vspace{1mm}
\noindent
\textit{Ablation Study:}
An ablation study (Figure \ref{fig:ablation_study}) highlights the contributions of each component. 
Removing the dual-head component
resulted in a significant performance drop in performance, especially on PC-GITA (35.2 F1 score), demonstrating the role of the two heads in capturing task-specific patterns. 
The exclusion of Adaptive Layers led to reduced performance on both datasets, confirming their importance for domain adaptation. 
While CNN-Based Bottleneck Layers were critical for PC-GITA, their removal slightly improved EWA-DB results, suggesting their impact may vary across linguistic contexts.
Removing wavelet integration and contrastive learning also led to performance declines, demonstrating the value of these features in enhancing model robustness and discrimination.

\vspace{1mm}
\noindent
\textit{t-SNE Visualization:}
Figure \ref{fig:tsne_comparison} shows a t-SNE plot of DDK embeddings (obtained after attention pooling), with colors representing different datasets and shapes for Healthy/Parkinson's samples.
Both models share the same architecture and training settings, except that the dual-head architecture is absent in \ref{fig:tsne_comparison} (a) and present in (b). 
Without the dual-head (Figure \ref{fig:tsne_comparison}(a)), the embeddings form two distinct clusters corresponding to EWA-DB and PC-GITA, indicating that the model primarily learns language-specific features. 
In contrast, the dual-head architecture (Figure \ref{fig:tsne_comparison}(b)) produces a more unified feature space with a smooth transition across data points from both datasets, demonstrating better generalization and shared feature learning across the two languages, rather than focusing on dataset-specific characteristics.

Overall, our findings demonstrate that the proposed model architecture, integrating multiple specialized components, effectively addresses the challenges of cross-language generalization and robust PD detection from speech.

\section{Conclusions}
\label{sec:conclusions}

We proposed a novel model architecture for PD detection from speech, designed to improve cross-language generalization by integrating SSL features, wavelet transforms, task-specific classifiers, and domain adaptation techniques. 
Our experiments demonstrate that the model generalizes effectively across different languages and datasets, outperforming baseline models in cross-dataset evaluations.
These findings suggest that robust PD detection models can be developed to handle diverse linguistic contexts, aiding in early diagnosis and monitoring across varied populations.
The challenge of acquiring large, diverse datasets for PD detection highlights the critical need for models that generalize effectively across various languages and recording conditions.
Future work will focus on refining these models and extending their applicability to additional languages and speech tasks, further enhancing their robustness and clinical utility.

\bibliographystyle{IEEEtran}
\bibliography{bibliography}

\begin{thebibliography}{10}
\providecommand{\url}[1]{#1}
\csname url@samestyle\endcsname
\providecommand{\newblock}{\relax}
\providecommand{\bibinfo}[2]{#2}
\providecommand{\BIBentrySTDinterwordspacing}{\spaceskip=0pt\relax}
\providecommand{\BIBentryALTinterwordstretchfactor}{4}
\providecommand{\BIBentryALTinterwordspacing}{\spaceskip=\fontdimen2\font plus
\BIBentryALTinterwordstretchfactor\fontdimen3\font minus \fontdimen4\font\relax}
\providecommand{\BIBforeignlanguage}[2]{{%
\expandafter\ifx\csname l@#1\endcsname\relax
\typeout{** WARNING: IEEEtran.bst: No hyphenation pattern has been}%
\typeout{** loaded for the language `#1'. Using the pattern for}%
\typeout{** the default language instead.}%
\else
\language=\csname l@#1\endcsname
\fi
#2}}
\providecommand{\BIBdecl}{\relax}
\BIBdecl

\bibitem{tsanas2012novel}
A.~Tsanas, M.~A. Little, P.~E. McSharry, J.~Spielman, and L.~O. Ramig, ``Novel speech signal processing algorithms for high-accuracy classification of parkinson's disease,'' \emph{IEEE transactions on biomedical engineering}, vol.~59, no.~5, pp. 1264--1271, 2012.

\bibitem{orozco2016automatic}
J.~R. Orozco-Arroyave, F.~H{\"o}nig, J.~Arias-Londo{\~n}o, J.~Vargas-Bonilla, K.~Daqrouq, S.~Skodda, J.~Rusz, and E.~N{\"o}th, ``Automatic detection of parkinson's disease in running speech spoken in three different languages,'' \emph{The Journal of the Acoustical Society of America}, vol. 139, no.~1, pp. 481--500, 2016.

\bibitem{Vasquez2018}
J.~Vásquez-Correa \emph{et~al.}, ``{Towards an Automatic Evaluation of the Dysarthria Level of Patients with Parkinson’s Disease},'' \emph{Journal of Communication Disorders}, vol.~76, pp. 21--36, 2018.

\bibitem{perez2019natural}
P.~A. P{\'e}rez-Toro \emph{et~al.}, ``Natural language analysis to detect {P}arkinson’s disease,'' in \emph{International Conference on Text, Speech, and Dialogue}.\hskip 1em plus 0.5em minus 0.4em\relax Springer, 2019, pp. 82--90.

\bibitem{vasquez2021transfer}
J.~V{\'a}squez-Correa \emph{et~al.}, ``Transfer learning helps to improve the accuracy to classify patients with different speech disorders in different languages,'' \emph{Pattern Recognition Letters}, vol. 150, pp. 272--279, 2021.

\bibitem{sonawane2021speech}
B.~Sonawane and P.~Sharma, ``Speech-based solution to {P}arkinson’s disease management,'' \emph{Multimedia Tools and Applications}, vol.~80, no.~19, pp. 29\,437--29\,451, 2021.

\bibitem{Narendra2021}
N.~Narendra, B.~Schuller, and P.~Alku, ``The detection of parkinson's disease from speech using voice source information,'' \emph{IEEE/ACM Transactions on Audio, Speech, and Language Processing}, vol.~29, pp. 1925--1936, 2021.

\bibitem{garcia2022detecting}
A.~Garc{\'\i}a \emph{et~al.}, ``Detecting {P}arkinson’s disease and its cognitive phenotypes via automated semantic analyses of action stories,'' \emph{{NPJ Parkinson's Disease}}, vol.~8, no.~1, pp. 163--10, 2022.

\bibitem{quan2022end}
C.~Quan \emph{et~al.}, ``End-to-end deep learning approach for {P}arkinson’s disease detection from speech signals,'' \emph{Biocybernetics and Biomedical Engineering}, vol.~42, no.~2, pp. 556--574, 2022.

\bibitem{favaro2023multilingual}
A.~Favaro, L.~Moro-Vel{\'a}zquez, A.~Butala, C.~Motley, T.~Cao, R.~D. Stevens, J.~Villalba, and N.~Dehak, ``Multilingual evaluation of interpretable biomarkers to represent language and speech patterns in parkinson's disease,'' \emph{Frontiers in Neurology}, vol.~14, p. 1142642, 2023.

\bibitem{Reddy2023}
M.~K. Reddy and P.~Alku, ``Exemplar-based sparse representations for detection of parkinson's disease from speech,'' \emph{IEEE/ACM Transactions on Audio, Speech, and Language Processing}, vol.~31, pp. 1386--1396, 2023.

\bibitem{bocklet2013automatic}
T.~Bocklet, S.~Steidl, E.~N{\"o}th, and S.~Skodda, ``Automatic evaluation of parkinson's speech-acoustic, prosodic and voice related cues.'' in \emph{Interspeech}.\hskip 1em plus 0.5em minus 0.4em\relax Citeseer, 2013, pp. 1149--1153.

\bibitem{la2024exploiting}
M.~{La Quatra}, M.~F. Turco, T.~Svendsen, G.~Salvi, J.~R. Orozco-Arroyave, and S.~M. Siniscalchi, ``Exploiting foundation models and speech enhancement for parkinson's disease detection from speech in real-world operative conditions,'' in \emph{Interspeech}, 2024.

\bibitem{escobargrisales23_interspeech}
D.~Escobar-Grisales, T.~Arias-Vergara, C.~D. Ríos-Urrego, E.~Nöth, A.~M. García, and J.~R. Orozco-Arroyave, ``An automatic multimodal approach to analyze linguistic and acoustic cues on parkinson's disease patients,'' in \emph{INTERSPEECH 2023}, 2023, pp. 1703--1707.

\bibitem{kovac2021multilingual}
D.~Kovac, J.~Mekyska, Z.~Galaz, L.~Brabenec, M.~Kostalova, S.~Z. Rapcsak, and I.~Rektorova, ``Multilingual analysis of speech and voice disorders in patients with parkinson’s disease,'' in \emph{2021 44th International Conference on Telecommunications and Signal Processing (TSP)}.\hskip 1em plus 0.5em minus 0.4em\relax IEEE, 2021, pp. 273--277.

\bibitem{baevski2020wav2vec}
A.~Baevski, Y.~Zhou, A.~Mohamed, and M.~Auli, ``wav2vec 2.0: A framework for self-supervised learning of speech representations,'' \emph{Advances in neural information processing systems}, vol.~33, pp. 12\,449--12\,460, 2020.

\bibitem{chen2022wavlm}
S.~Chen, C.~Wang, Z.~Chen, Y.~Wu, S.~Liu, Z.~Chen, J.~Li, N.~Kanda, T.~Yoshioka, X.~Xiao \emph{et~al.}, ``Wavlm: Large-scale self-supervised pre-training for full stack speech processing,'' \emph{IEEE Journal of Selected Topics in Signal Processing}, vol.~16, no.~6, pp. 1505--1518, 2022.

\bibitem{hsu2021hubert}
W.-N. Hsu, B.~Bolte, Y.-H.~H. Tsai, K.~Lakhotia, R.~Salakhutdinov, and A.~Mohamed, ``Hubert: Self-supervised speech representation learning by masked prediction of hidden units,'' \emph{IEEE/ACM transactions on audio, speech, and language processing}, vol.~29, pp. 3451--3460, 2021.

\bibitem{Mallat2008}
S.~Mallat, \emph{A Wavelet Tour of Signal Processing}, 3rd~ed.\hskip 1em plus 0.5em minus 0.4em\relax USA: Academic Press, Inc., 2008.

\bibitem{huang2017adain}
X.~Huang and S.~Belongie, ``Arbitrary style transfer in real-time with adaptive instance normalization,'' in \emph{Proceedings of the IEEE international conference on computer vision}, 2017, pp. 1501--1510.

\bibitem{rusko2023ewa}
M.~Rusko, R.~Sabo, M.~Trnka, A.~Zimmermann, R.~Malaschitz, E.~Ru{\v{z}}ick{\`y}, P.~Brandoburov{\'a}, V.~Kevick{\'a}, and M.~{\v{S}}korv{\'a}nek, ``Ewa-db, slovak database of speech affected by neurodegenerative diseases,'' \emph{medRxiv}, pp. 2023--10, 2023.

\bibitem{orozco2014new}
J.~R. Orozco-Arroyave, J.~D. Arias-Londo{\~n}o, J.~F. Vargas-Bonilla, M.~C. Gonzalez-R{\'a}tiva, and E.~N{\"o}th, ``New spanish speech corpus database for the analysis of people suffering from parkinson's disease.'' in \emph{Lrec}, 2014, pp. 342--347.

\bibitem{karan2021non}
B.~Karan, S.~S. Sahu, J.~R. Orozco-Arroyave, and K.~Mahto, ``Non-negative matrix factorization-based time-frequency feature extraction of voice signal for parkinson's disease prediction,'' \emph{Computer Speech \& Language}, vol.~69, p. 101216, 2021.

\bibitem{ibarra2023towards}
E.~J. Ibarra, J.~D. Arias-Londo{\~n}o, M.~Za{\~n}artu, and J.~I. Godino-Llorente, ``Towards a corpus (and language)-independent screening of parkinson’s disease from voice and speech through domain adaptation,'' \emph{Bioengineering}, vol.~10, no.~11, p. 1316, 2023.

\bibitem{woszczyk20_interspeech}
D.~Woszczyk, S.~Petridis, and D.~Millard, ``Domain adversarial neural networks for dysarthric speech recognition,'' in \emph{Interspeech 2020}, 2020, pp. 3875--3879.

\bibitem{wang21u_interspeech}
D.~Wang, L.~Deng, Y.~T. Yeung, X.~Chen, X.~Liu, and H.~Meng, ``Unsupervised domain adaptation for dysarthric speech detection via domain adversarial training and mutual information minimization,'' in \emph{Interspeech 2021}, 2021, pp. 2956--2960.

\bibitem{rios2024automatic}
C.~D. Rios-Urrego, J.~Rusz, and J.~R. Orozco-Arroyave, ``Automatic speech-based assessment to discriminate parkinson’s disease from essential tremor with a cross-language approach,'' \emph{npj Digital Medicine}, vol.~7, no.~1, p.~37, 2024.

\bibitem{Orozco2016}
J.~R. Orozco-Arroyave, F.~T. Hönig, J.~D. Arias-Londoño, J.~F. Vargas-Bonilla, K.~Daqrouq, S.~Skodda, J.~Rusz, and E.~Nöth, ``{Automatic} detection of {Parkinson}'s disease in running speech spoken in three different languages.'' \emph{The Journal of the Acoustical Society of America}, vol. 139, p. 481, Jan 2016.

\bibitem{torghabeh2023enhancing}
F.~A. Torghabeh, S.~A. Hosseini, and E.~A. Moghadam, ``Enhancing parkinson's disease severity assessment through voice-based wavelet scattering, optimized model selection, and weighted majority voting,'' \emph{Medicine in Novel Technology and Devices}, vol.~20, p. 100266, 2023.

\bibitem{tan2020rvad}
Z.-H. Tan, N.~Dehak \emph{et~al.}, ``rvad: An unsupervised segment-based robust voice activity detection method,'' \emph{Computer speech \& language}, vol.~59, pp. 1--21, 2020.

\bibitem{welker2022speech}
S.~Welker, J.~Richter, and T.~Gerkmann, ``Speech enhancement with score-based generative models in the complex stft domain,'' \emph{arXiv preprint arXiv:2203.17004}, 2022.

\bibitem{lu23e_interspeech}
Y.-X. Lu, Y.~Ai, and Z.-H. Ling, ``Mp-senet: A speech enhancement model with parallel denoising of magnitude and phase spectra,'' in \emph{INTERSPEECH 2023}, 2023, pp. 3834--3838.

\bibitem{loshchilov2018decoupled}
I.~Loshchilov and F.~Hutter, ``Decoupled weight decay regularization,'' in \emph{International Conference on Learning Representations}, 2019.

\end{thebibliography}

\end{document}